\documentclass[reprint,superscriptaddress,aps,reprint,showpacs]{revtex4-2}
\usepackage{graphicx}
\usepackage{dcolumn}
\usepackage[dvipsnames]{xcolor}
\usepackage{bm}
\usepackage{color}
\usepackage{scrextend}		
\usepackage{url}
\usepackage{epstopdf}
\usepackage{hyperref}
\usepackage{placeins}
\usepackage{siunitx} 
\pdfoutput=1

\begin{document}

\title{Unraveling the origin of antiferromagnetic coupling at YIG/permalloy interface}

\author{Jiangchao Qian}
\affiliation{Materials Research Laboratory and Department of Materials Science and Engineering, University of Illinois at Urbana-Champaign, Urbana, Illinois 61801, USA}
\author{Yi Li}
\affiliation{Materials Science Division, Argonne National Laboratory, Argonne, IL 60439, USA}
\author{Zhihao Jiang}
\affiliation{Materials Research Laboratory and Department of Materials Science and Engineering, University of Illinois at Urbana-Champaign, Urbana, Illinois 61801, USA}
\author{Robert Busch}
\affiliation{Materials Research Laboratory and Department of Materials Science and Engineering, University of Illinois at Urbana-Champaign, Urbana, Illinois 61801, USA}
\author{Hsu-Chih Ni}
\affiliation{Materials Research Laboratory and Department of Materials Science and Engineering, University of Illinois at Urbana-Champaign, Urbana, Illinois 61801, USA}
\author{Tzu-Hsiang Lo}
\affiliation{Materials Research Laboratory and Department of Materials Science and Engineering, University of Illinois at Urbana-Champaign, Urbana, Illinois 61801, USA}
\author{Axel Hoffmann}
\affiliation{Materials Research Laboratory and Department of Materials Science and Engineering, University of Illinois at Urbana-Champaign, Urbana, Illinois 61801, USA}
\author{Andr\'e Schleife}
\affiliation{Materials Research Laboratory and Department of Materials Science and Engineering, University of Illinois at Urbana-Champaign, Urbana, Illinois 61801, USA}
\author{Jian-Min Zuo}
\email{Electronic mail: jianzuo@illinois.edu}
\affiliation{Materials Research Laboratory and Department of Materials Science and Engineering, University of Illinois at Urbana-Champaign, Urbana, Illinois 61801, USA}
\email{Electronic mail: jianzuo@illinois.edu}

\date{\today}

\begin{abstract}

We investigate the structural and electronic origin of antiferromagnetic (AFM) coupling in the Yttrium iron garnet (YIG) and permalloy ($\text{Ni}_{80}\text{Fe}_{20}$, Py) bilayer system at the atomic level. Ferromagnetic Resonance (FMR) reveal unique hybrid modes in samples prepared with surface ion milling, indicative of antiferromagnetic exchange coupling at the YIG/Py interface. Using atomic resolution scanning transmission electron microscopy (STEM), we found that AFM coupling appears at the YIG/Py interface of the \textit{tetrahedral} YIG surface formed with ion milling. The STEM measurements suggest that the interfacial AFM coupling is predominantly driven by an oxygen-mediated super-exchange coupling mechanism, which is confirmed by the density functional theory (DFT) calculations to be energetically favorable. Thus, the combined experimental and theoretical results reveal the critical role of interfacial atomic structure in determining the type magnetic coupling in a YIG/ferromagnet heterostructure, and prove that the interfacial structure can be experimentally tuned by surface ion-milling.


\end{abstract}
\maketitle


\section{\label{sec:level1}Introduction}

Yttrium iron garnet (Y$_{3}$Fe$_{5}$O$_{12}$) is well-known for its low magnetic damping \cite{word1,word2,NEW1_Hauser2016,Fengyuan2017,Fengyuan2022,Fengyuan2023,Takian2024}, making it the material of choice for efficient spin interactions in magnonics \cite{NEW2_Serga_2010,NEW3_Yu2014,NEW4_collet2016,NEW5_APL2023}, spin transport \cite{NEW6_Cornelissen2015}, cavity spintronics \cite{NEW7_PRL_2013,NEW8_PRL_2014,NEW9_PRL_2014Zhang,NEW10_PRL2015}, and quantum information science \cite{NEW11_sci2020,NEW12_PRL2023}. Considerable attention has been directed toward building YIG-based thin film heterostructures for spin-based information processing by taking advantage of interfacial spin interactions. One example is given by YIG/Pt bilayers, with experimental observations of spin pumping \cite{NEW13_PRB2015}, nonlocal spin injection \cite{NEW6_Cornelissen2015}, and spin Hall magnetoresistance \cite{NEW14_PRL2013}, where the interlayer exchange coupling has significantly enhanced the spin transmission across the YIG-Pt interface.

Recently, another YIG heterostructure with a ferromagnetic (FM) layer, i.e, YIG/FM bilayer, has aroused increasing interests owing to its potential in hybrid magnonics from the interlayer magnon-magnon coupling \cite{Stefan2018PRL,NEW16_PRL2018,NEW17_Cornelissen2015,Yi2020PRL}. The ferromagnetic resonance mode in the FM layer can form strong coupling with the perpendicular standing spin wave modes in YIG due to the interfacial exchange interaction, leading to new physical phenomena such as coherent spin pumping \cite{Yi2020PRL}, magnetically induced transparency \cite{NEW20_Xiong2022}, and efficient excitations of short-wavelength spin waves \cite{NEW21_Inman2022}. However, the physical mechanisms underlying the interfacial exchange coupling, crucial for coupling spin excitations between the two magnetic systems, remain not fully understood. Particularly, recent works have revealed pronounced antiferromagnetic coupling across YIG/CoFeB, YIG/Co, and YIG/$\text{Ni}_{80}\text{Fe}_{20}$~(Py) interfaces, where the origin of antiferromagnetic coupling between YIG and FM layers has been a topic of considerable debate. Previous studies, reported either ferromagnetic coupling or antiferromagnetic coupling in different prepared YIG/FM bilayers \cite{Luqiao2020PRApplied,Luqiao2022PRM,Stefan2018PRL}. Specifically, Fan et al. \cite{Luqiao2020PRApplied}, Quarterman et al. \cite{Luqiao2022PRM}, and Klingler et al. \cite{Stefan2018PRL} have posited theories of direct exchange coupling, while the possibility of super-exchange coupling also remains. These findings hint at the significant role of interfacial structure in determining bilayer coupling mechanisms, yet tuning magnetic interactions through interface engineering remains insufficiently explored. Here we elucidate the interface structure between YIG and Py and their magnetic coupling. By integrating FMR with STEM/EELS and DFT calculations, our study reveals the significant role of surface treatments using ion-milling in enhancing antiferromagnetic coupling through promoting the oxygen mediated super-exchange coupling mechanisms at the YIG/Py interface. 

\section{\label{sec:level1}Methods}

\begin{figure}[htb]    
 \centering
 \includegraphics[width=3.25 in, angle = 0]{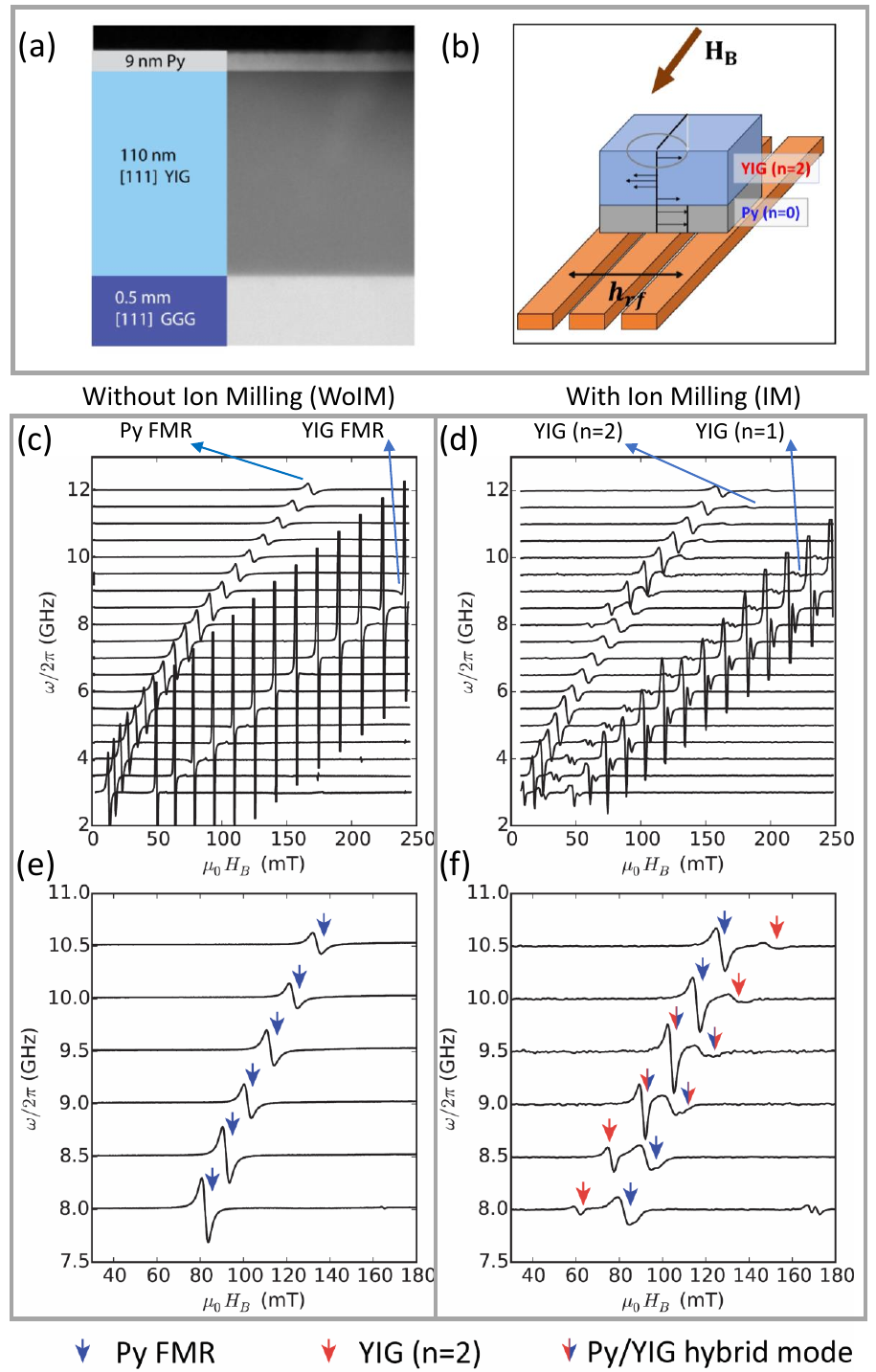}
 \caption{The YIG/Py bilayer structure and FMR characterization. (a) STEM image together with the designed bilayer structure. (b) Illustration of the FMR setup: The coplanar waveguide is located underneath the YIG/Py bilayer thin films. (c)-(d): Full-range FMR signals of the YIG/Py (c) without ion-milling (WoIM) and (d) with ion-milling samples (IM) samples measured by the coplanar waveguide. (e)-(f): Zoom-in view of (c)-(d) in the regime where mode hybridization between the Py FMR mode and the YIG (n=2) PSSW mode is observed. }
 \label{FIG-FMR}
\end{figure} 

In line with the preparation of YIG/Py bilayer in our prior work \cite{Yi2020PRL}, we first deposited YIG (100 nm) onto two (111)-oriented  Gd$_{3}$Ga$_{5}$O$_{12}$ substrates by magnetron sputtering. Then the amorphous YIG films were annealed in air at 850 °C for 3 hours, and slowly cooled down to room temperature by 0.5 °C/min, yielding epitaxial YIG films with light yellow color. To study the formation of antiferromagnetic interfacial exchange coupling, we slightly ion milled one YIG film in the sputtering chamber under Ar environment by applying an RF bias voltage through the substrate holder, where the holder acts as an effective sputtering gun, and trigger Ar$^{+}$ ion bombarding of the substrate surface. The milling rate was 3 nm/mins and the milling process lasted for 1 min 30 s. A Py (10 nm) thin film was subsequently sputtered on the milled YIG film without breaking the vacuum, ensuring efficient interfacial exchange coupling. A control YIG/Py bilayer sample was also prepared without ion milling the YIG surface. Figure~\ref{FIG-FMR}a shows the cross-sectional film structure. The quality of the bilayers was checked using X-Ray Diffraction (XRD) and the result indicates a much sharper YIG/Py interface as the result of ion-milling \cite{SI}.




\section{\label{sec:level1}Results}


In order to study the interfacial exchange coupling, we first conducted FMR measurements on the two YIG/Py samples. Figure 1b details this experimental arrangement with the samples flipped on top of a coplanar waveguide for microwave transmission measurements \cite{Yi2020PRL}. By modulating the biasing field, the transmitted microwave is converted to a DC voltage using a tunnel diode, and then measured by a lock-in amplifier. Figs.~\ref{FIG-FMR}(c-f) show the measured signals, which correspond to the derivative FMR signals. As can be seen, the YIG and Py FMR signals can be clearly measured in both samples, showing good quality of the YIG and Py films. Compared with Py, the YIG resonance signals are located at higher fields and show stronger and narrower lines due to a smaller demagnetizing field and a much lower magnetic damping of YIG. From the FMR data, the existence of interfacial exchange coupling in the IM sample is supported by three observations. First, the YIG linewidth becomes broader because the interfacial exchange coupling provide an additional contribution to spin pumping from YIG to Py. Second, perpendicular standing spin wave modes of YIG (n=1 and n=2) can be measured because the interfacial exchange coupling breaks the symmetry of the standing wave and leads to a finite coupling to the uniform microwave field of the coplanar waveguide. Third, clear mode hybridization is measured between the Py FMR mode and the YIG (n=2) standing spin wave modes at around 9 GHz, which is emphasized in Fig.~\ref{FIG-FMR}(f). Note that the lower-field hybrid mode between 9 and 10 GHz shows a narrower linewidth compared with the higher-field hybrid mode. This means the lower- (higher-) field mode corresponds to the in-phase (out-of-phase) dynamics because of the absence (existence) of coherent spin pumping between YIG and Py \cite{Yi2020PRL}. One can thus deduce that the interfacial exchange coupling is antiferromagnetic because out-of-phase dynamics softens the hybrid dynamics and a stronger field is needed in order to maintain the resonance at the same frequency.

\begin{figure}[htb]  
 \centering
 \includegraphics[width= 3.2 in, angle = 0]{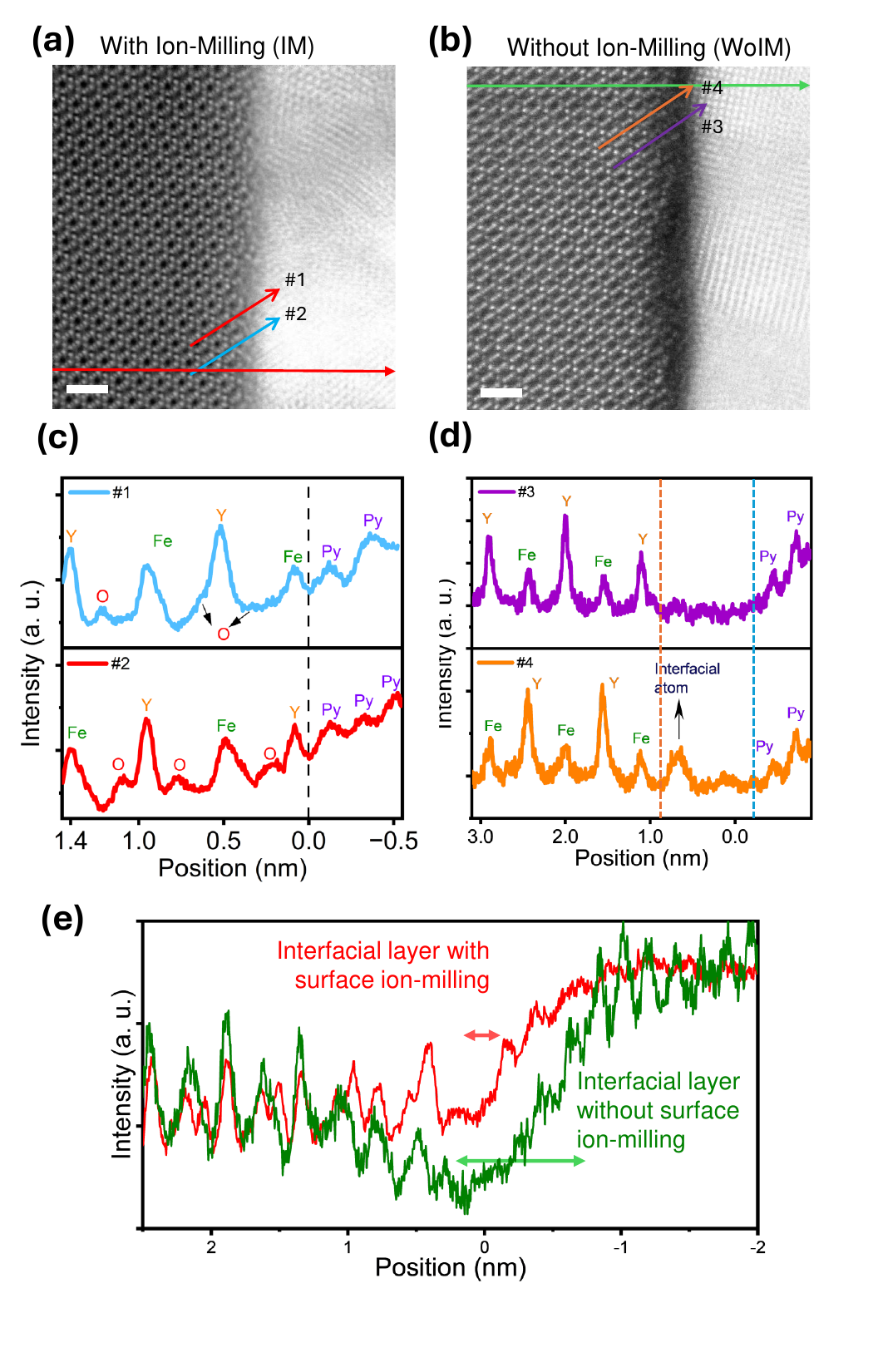}
 \caption{Interfacial structure as seen by atomic-resolution STEM-HAADF along the YIG [110] zone axis. STEM-HAADF images for the with ion-milling (a) and without ion-milling (b) samples. The scale bars are both 0.5 nm. (c,d) The line profiles of HAADF intensity across the the marked lines 1 to 4 in (a,b), showing the termination of Y and Fe, with Fe/Ni atoms from Py layer adjacent to YIG. (d) The line profiles of HAADF intensity across the arrows lines in (b), showing the termination of Y and Fe, with an interfacial layer adjacent to YIG. The interfacial layer is about 1 nm width. (e) The marked horizontal line profiles of HAADF images in (a) and (b) show the different YIG/Py interface width.}
 \label{FIG-HAADF110}
\end{figure} 

To investigate the structural origin of the magnetic differences between samples with and without ion-milling, we performed cross-sectional STEM of the YIG/Py bilayer films using a high-angle annular dark-field detector (HAADF) for Z-contrast \cite{SI}. Figure~\ref{FIG-HAADF110}a and~\ref{FIG-HAADF110}b  show the STEM-HAADF images of the two samples, providing an atomic-scale inspection of their interfacial structural differences. The YIG surface termination in the IM sample, as revealed in Fig.~\ref{FIG-HAADF110}a, comprises a plane of visible Fe and Y atoms, and invisible O atoms, leading to an interface layer that sharply connects to the polycrystalline Py. In contrast, the interface in the WoIM sample exhibits roughness and features an amorphous layer approximately 0.7 nm in width (Fig.~\ref{FIG-HAADF110}b). The small contrast difference of the YIG film in Fig.~\ref{FIG-HAADF110}b is due to a change in thickness and sample orientation. The direct adjacency of the YIG termination plane to the Py layer in the IM sample eliminates the observed dark contrast layer in the WoIM sample. This distinction in interfacial structure between the two samples is further substantiated by the intensity line profiles depicted in Figs.~\ref{FIG-HAADF110}c,d,e, taken along the lines marked in ~\ref{FIG-HAADF110}a,b. These profiles demonstrate that the termination plane ends with a combination of Y and Fe containing atomic columns, with Ni/Fe atoms from Py side directly adjoining the crystalline garnet structure in the IM sample. 
In the WoIM sample, the intensity profiles indicate the presence of an interfacial layer that measures about 0.7 nm in width. The surface ion-milling process prior to the deposition of Py removes this interfacial layer in the IM sample.


\begin{figure}[!]  
 \centering
 \includegraphics[width= 3.2 in, angle = 0]{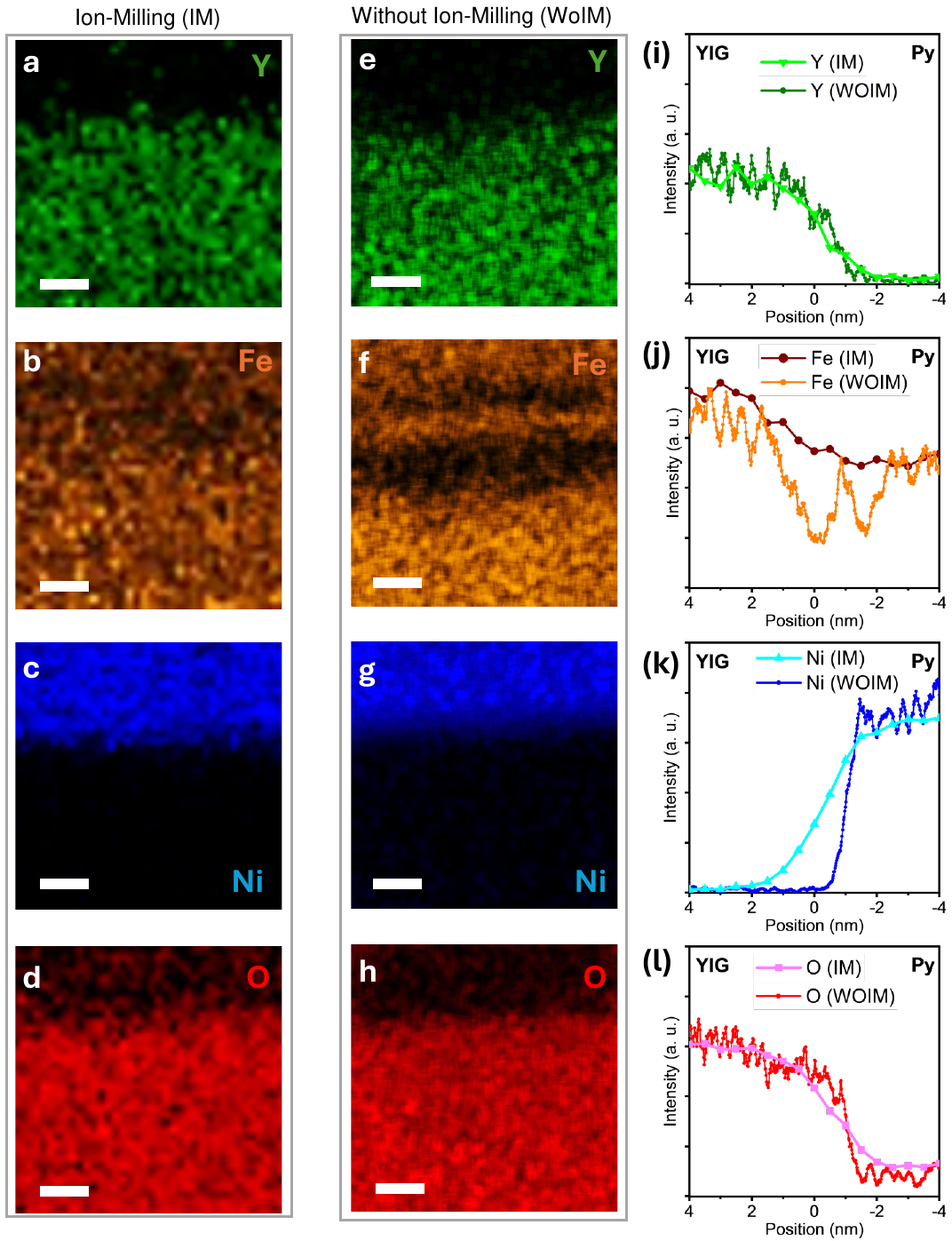}
 \caption{
 EDS elemental analysis for the samples with and without surface ion-milling. (a-d): Elemental maps of Y, Fe, Ni, and O for the sample with surface ion-milling (IM). (e-h):  Elemental maps of Y, Fe, Ni, and O for the sample without surface ion-milling (WoIM). The scale bars are 2 nm. (i-j) Line profiles comparing elemental distributions for Y, Fe, Ni, and O across the YIG/Py interfaces in IM and WoIM samples. 
  }
 
 \label{FIG-EDS}
\end{figure}

Figure~\ref{FIG-EDS} compares the chemical distribution between two samples, using the energy-dispersive X-ray spectroscopy (EDS) mapping performed in the STEM mode at the pixel size of about 1 nm. The main differences between the IM and WoIM samples are Fe deficiency and Ni diffusion near the YIG surface. This surface Fe deficiency of the as grown YIG surface has been previously reported by Sun et al. \cite{Sun_Axel_Fe_loss}. The EDS mappings show that ion-milling effectively removes the Fe-deficient layer and provides a direct transition between the YIG and Py layers. The EDS mapping also indicates that Ni penetrates into the ion-milled YIG, suggesting an opening of diffusion channels with ion-milling. Overall, the benefit of ion-milling is to remove the reconstructed YIG surface caused by annealing.


\begin{figure*}  
\centering
\includegraphics[width= 6 in, angle = 0]{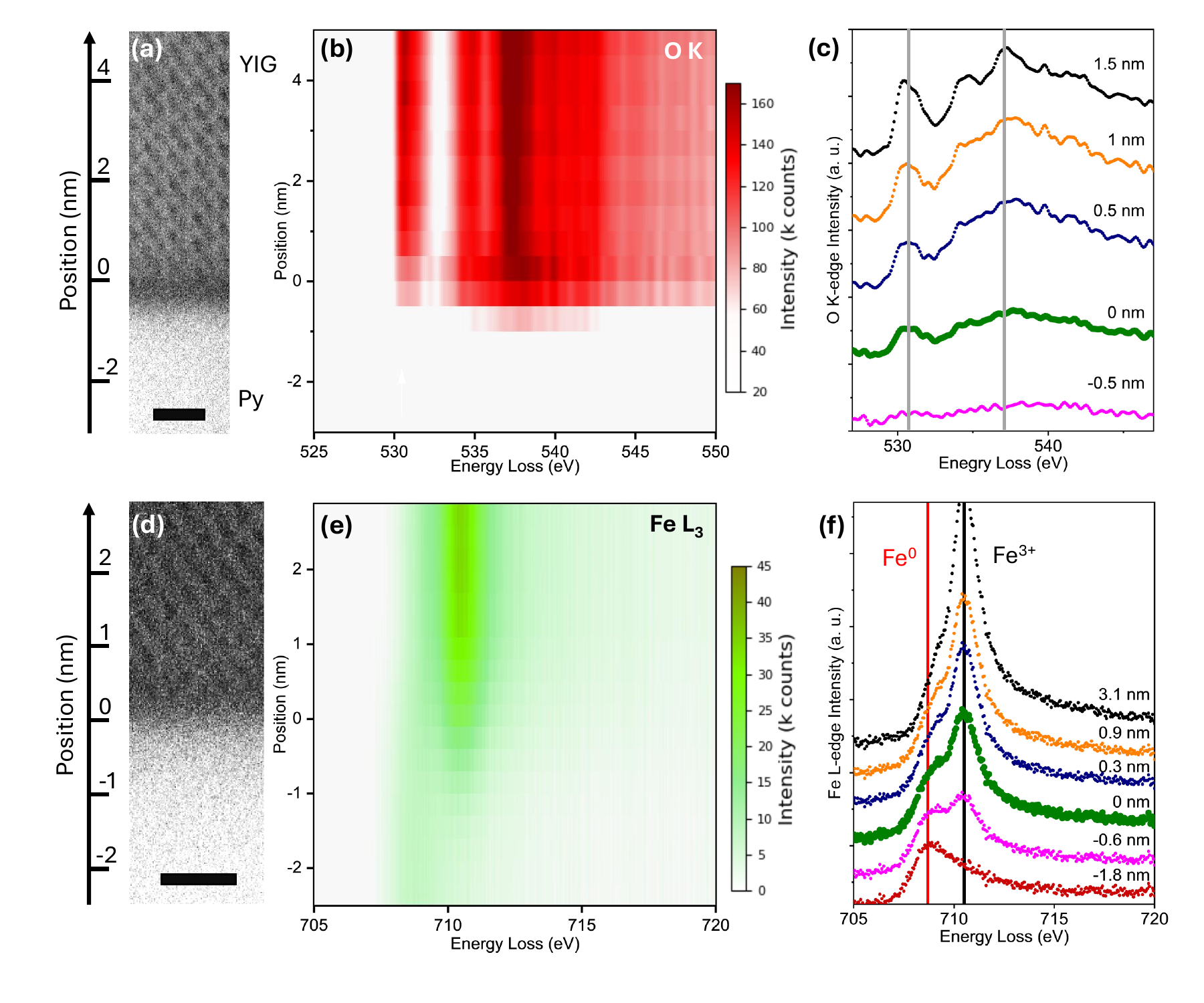}
\caption{
The EELS near-edge fine structure analysis with corresponding HAADF images of the YIG/Py interface in sample IM. (a) and (d): HAADF images showing the EELS O K-edge, Fe $L_{3}$-edge survey region for the surface ion-milled sample, respectively. The scale bars are 1 nm. (b) and (e): Integrated EELS spectra horizontally with 0 nm position set at the YIG/Py interfaces. (c) and (f): Hyperspectral images for O K-edge (c) and (f) Fe $L_{3}$-edge, respectively, illustrating the evolution of  the fine structures. The results here directly evident of the presence of O$^{2-}$ and Fe$^{3+}$ at the interface.
}
\label{FIG-EELS}
\end{figure*} 

From our STEM-HAADF observations, the garnet structure appears consistent up to the interface, suggesting a minimal structural modification. Electron energy loss spectroscopy (EELS) analysis was performed in the STEM mode to further examine the chemical sharpness of the YIG/Py interface in the IM sample. Figures~\ref{FIG-EELS}b,c display the oxygen K-edge fine structure, alongside a HAADF survey image (Fig.~\ref{FIG-EELS}a) with a sampling resolution of 0.5 nm. This combination provides insights into both the oxygen content and the nature of its chemical bonding. The oxygen K-edge fine features are observed up to the interface. A drop in the K-edge intensity is observed near within the distance of ~1 nm to the interface, which can be attributed to the electron probe spreading effect \cite{ShahMicron}. In addition to the pre-peak for oxygen, the fine features in the range of 535 eV to 540 eV also exhibit changes near the interfaces, indicative of slightly altered yttrium-oxygen (Y-O) bonding. Figures~\ref{FIG-EELS}e,f showcase the EELS fine structures for the Fe L$_{3}$-edge, coupled with a HAADF survey image (Fig.~\ref{FIG-EELS}d). The marked peaks of $708.7$ eV and $710.5$ eV in Fig.~\ref{FIG-EELS}f are features for Fe$^{0}$ and Fe$^{3+}$ respectively. The lower layer consists of Fe$^{0}$  from Py, while the upper layer contains Fe$^{3+}$  from YIG. The evolution of the oxygen K-edge and Fe L$_{3}$ edge are further highlighted in Fig.~\ref{FIG-EELS}c,f, which plot the hyperspectral EELS data versus the STEM probe position. At the interface, the Fe L$_{3}$ EELS signal is approximately a composite of Fe$^{0}$ and Fe$^{3+}$, while O$^{2-}$ is detected up to the interface. The presence of O$^{2-}$ and Fe$^{3+}$ at the interface and slightly beyond, and its sharper transition than the L$_{3}$ edge of Fe, suggests an oxygen-terminated YIG surface. The fine feature of L$_{3}$ edge shows the dominant peak is still $710.5$ eV with a wider shoulder left.The coexistence of Fe$^{0}$ and Fe$^{3+}$ in the interfacial transition region suggests a small amount of interfacial defects, possibly Fe interstitial atoms inside a rather open YIG structure. The EELS map shows a oxidized YIG surface with remaining garnet structures directly adjacent to the Py layer.

\begin{figure*} [hbt]
\centering
\includegraphics[width=5 in, angle = 0]{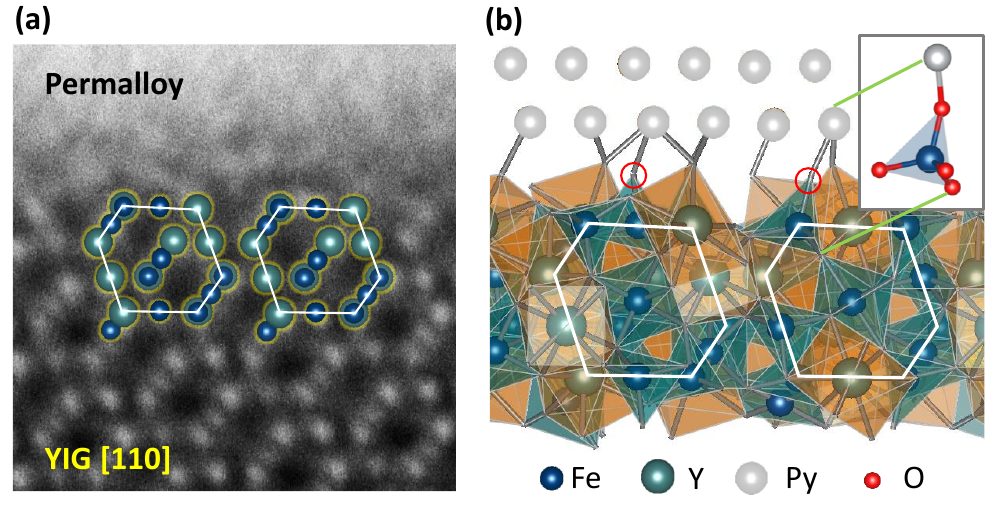}
\caption{YIG surface termination and structure model as viewed along [110] in sample IM. (a) Zoomed-in HAADF image in Fig.~\ref{FIG-HAADF110}a showing the overlap with YIG motif structures at surface terminations. (b) YIG/Py slab model for Density Functional Theory (DFT) simulation. Highlighted white hexagons represent YIG motifs. The inset illustrates the coupling between Py atom and \textit{tetrahedral} coordinated Fe}
\label{FIG-Model}
\end{figure*}

\section{\label{sec:level1}Discussion}

\begin{figure*} [hbt] 
    \centering
    \includegraphics[width= 6.5 in, angle = 0]{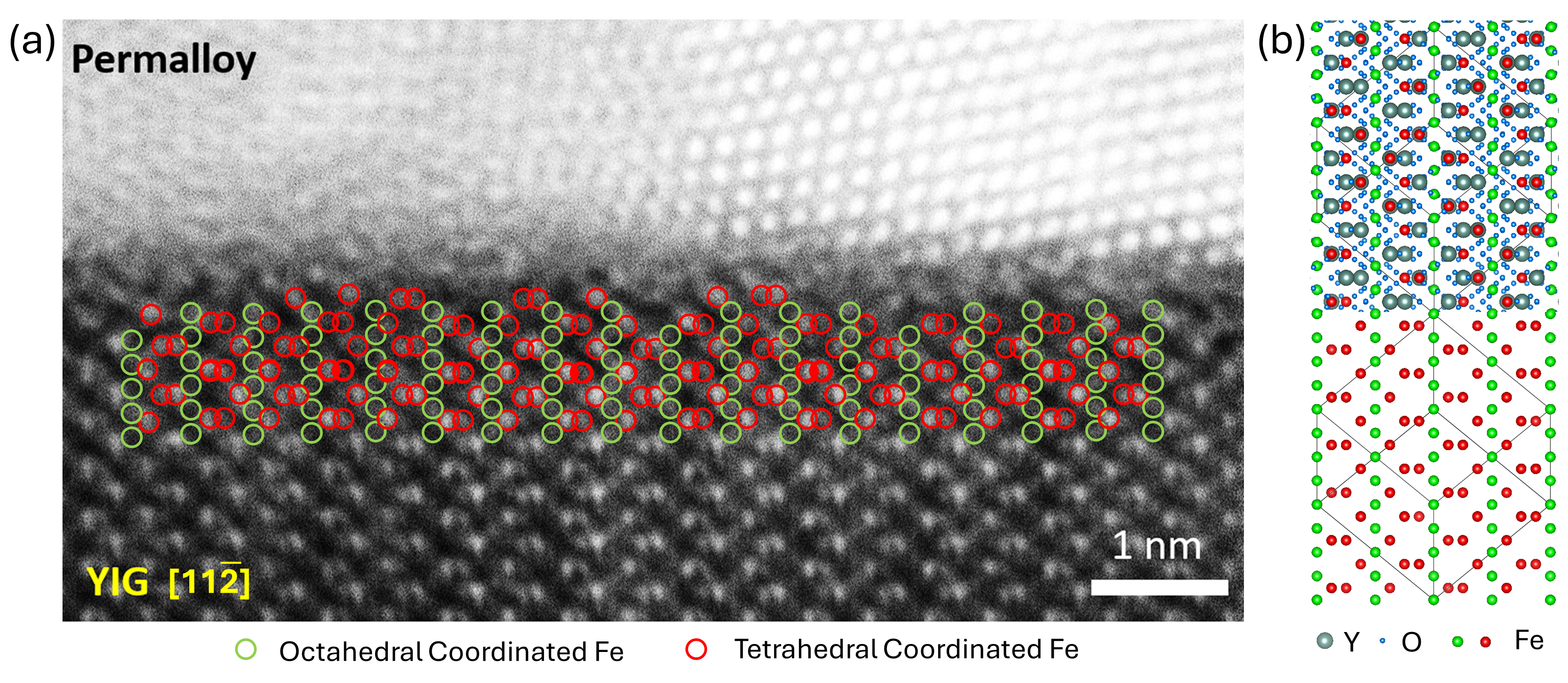}
    \caption{
    YIG surface termination and structure model as viewed along [11-2] in sample IM. (a) Atomic-resolution cross-sectional STEM-HAADF images of the YIG/Py interface. The YIG surface has a mixture of \textit{tetrahedral-} and \textit{octahedral-} coordinated Fe (marked as red and green, respectively) and Y atoms (not shown). The \textit{tetrahedral-} coordinated Fe prefers AFM coupling, while the \textit{octahedral-} coordinated Fe can be AFM or FM coupled according to DFT. The permalloy layer at the interface appears disordered and polycrystalline further away from the interface. The scale bar is 1 nm. (b) The YIG structure model from the [11-2] projection with (top) and without (bottom) Y and O atoms. 
    }
    
    \label{FIG-HAADF112}
\end{figure*}

To uncover the origin of experimentally observed antiferromagnetic coupling at the ion-milled interface of YIG/Py theoretically, we employed density functional theory (DFT) calculations to simulate the YIG/Py interfacial configurations. The interface structural models we proposed based on the experimental observations are depicted in Figs.~\ref{FIG-Model}a,b and Fig.~\ref{FIG-HAADF112}. The YIG (111) surface is terminated with yttrium, coordinated by 8 oxygen ions, and \textit{tetrahedral} coordinated Fe, respectively. The superposition of the YIG structural motifs on the magnified atomic resolution image from Fig.~\ref{FIG-Model}a shows good agreement between the model and the observed interfacial atomic arrangements in YIG. The interfacial Py atoms appear disordered and for simplicity, we have assumed two perfect layers of Py atoms parallel to the YIG surface in our model.  

The sharpness of the ion-milled YIG/Py interface was further examined from the $[11\bar{2}]$ orientation, as shown in Fig.~\ref{FIG-HAADF112}. From this orientation, the \textit{octahedral} Fe atoms are aligned in planes perpendicular to the horizontal interface. The YIG surface is seen to exhibit a mixture of \textit{tetrahedral} and \textit{octahedral} Fe terminations, with the \textit{tetrahedral} Fe terminations being the predominantly observed feature. The Fe/Ni atoms in the Py layer are polycrystalline and cannot be distinguished in the observation. Near the interface, the Fe/Ni atoms appear disordered due to interactions with the YIG surface, and are also not individually resolved in Fig.~\ref{FIG-Model}a.


Two potential magnetic arrangements can exist at the YIG/Py interface: antiferromagnetic (AFM), where bulk YIG and Py have opposite magnetic moments, and ferromagnetic with same magnetic moments. Using the structure model in Fig.~\ref{FIG-Model}b, we performed spin-polarized DFT calculation with different Fe and Ni arrangements in Py~\cite{SI}. The results indicate that the AFM arrangement is energetically more favorable than FM arrangement with an average energy difference of 2.2 meV per atom. 
However, when we placed the Py layers on the YIG slab with the surface termination of \textit{octahedral} coordinated Fe, the simulation results show no preference between FM and AFM interficial coupling~\cite{SI}.
Within the YIG, \textit{octahedral} Fe and \textit{tetrahedral} Fe are antiferromagnetically coupled, while the \textit{tetrahedral} Fe provides the dominant spin. In the AFM arrangement, the spins of \textit{tetrahedral} Fe are also antiferromagnetically coupled with Py atoms. The DFT simulation results thus supports our experimental observation of AFM in the IM sample where the \textit{tetrahedral} Fe ions interacting with Py atoms with the mediation of oxygen ions (as highlighted by circles in Fig.~\ref{FIG-Model}b) contribute to AFM coupling. The surface on the YIG without the ion-milling in comparison is rough with a mixture of surface terminations in addition to the reconstructed surface with Fe deficiency that breaks the AFM coupling between bulk YIG and Py. 


We note that the AFM coupling is not specific to the YIG/Py system. Previous works by Klingler et al. \cite{Stefan2018PRL} have reported interfacial antiferromagnetic coupling in YIG/Co. In their experiment, the liquid-phase-epitaxy (LPE) grown YIG film is cleaned with Piranha Etch and then annealed in oxygen. The magnetic coupling possibly also depends on the orientation of the YIG film, however, this has yet to be studied systematically\cite{Luqiao2022PRM}. On the other hand, the strong AFM coupling in the (111) oriented YIG with Py has been observed by Li et al. using the same preparation procedures as we described in this work \cite{Yi2020PRL}. The (111) orientation of YIG studied here is particularly significant because of its minimal lattice parameter mismatch with the GGG substrate, which facilitates the epitaxial growth of high-quality thin films.  Additionally, the (111) oriented YIG surface is the most dense-packed.

Given the coexistence of (Ni/Fe)$^{0}$ , Fe$^{3+}$ , and oxygen at the YIG/Py interface, we propose that oxygen-mediated super-exchange coupling could be the predominant mechanism for the antiferromagnetic interaction between Py and YIG, conceptualized as $<(\text{Ni/Fe})^{0}|\uparrow>--$O$--<\text{Fe}^{3+}|\downarrow>$. The substitution of Ni$^{0}$  for Fe$^{0}$  likely does not alter the direction of the magnetic moments. The \textit{octahedral} Fe terminated YIG/Py interface can have either AFM or FM coupling without a particular preference. The combination of these factors favors AFM coupling. This insight is crucial in our discussion, as it suggests that the antiferromagnetic coupling between Py and YIG is predominantly driven by oxygen-mediated super-exchange coupling.

\section{\label{sec:level1}conclusion}
 
In conclusion, we have identified the interfacial atomic structure as a key factor in determining the type of magnetic coupling between YIG and Py. Our FMR measurements imply that AFM coupling is associated with a sharp YIG/Py interface.
Ion-milling effectively removes the Fe deficiency layer on the annealed YIG, which appears to be the main beneficial factor. Our STEM observation also identifies the termination of \textit{tetrahedral} Fe as the first magnetic layer on the YIG side, which is mixed with the rough regions from a stepped surface. 
DFT calculations confirm that AFM coupling is an energetically favorable magnetic state for \textit{tetrahedral} Fe-terminated YIG/Py, while the \textit{octahedral} Fe-terminated YIG/Py interface can exhibit either AFM or FM coupling without a particular preference.
The preference for AFM coupling in the \textit{tetrahedral} Fe-terminated interface structure is due to inter-layer super-exchange interaction mediated by the oxygen layer in-between. This mechanism explains the origin of AFM coupling as oxygen-mediated super-exchange interaction between \textit{tetrahedral} Fe in YIG and magnetic atoms (Ni/Fe) in Py.

\begin{acknowledgments}
The work was supported by the U.S.\ DOE, Office of Science, Basic Energy Sciences, Materials Sciences and Engineering Division, with all parts of the manuscript preparation supported under contract No.\ DE-SC0022060. This work made use of the Illinois Campus Cluster, a computing resource that is operated by the Illinois Campus Cluster Program (ICCP) in conjunction with the National Center for Supercomputing Applications (NCSA) and which is supported by funds from the University of Illinois at Urbana-Champaign.
\end{acknowledgments}

\clearpage 
\bibliographystyle{apsrev4-2} 
\bibliography{main}

\end{document}